%
%
\input harvmac.tex

\def\cn{{\cal N}}
\def\IR{\relax{\rm I\kern-.18em R}}
\def\IZ{\relax\ifmmode\hbox{Z\kern-.35em Z}\else{Z\kern-.35em Z}\fi}
\def\RP{{\bf RP}}

\def\bS{{\bf S}}

%
%
\def\np#1#2#3{{\it Nucl. Phys.} {\bf B#1} (#2) #3}

\def\plb#1#2#3{{\it Phys. Lett.} {\bf #1B} (#2) #3}

\def\prep#1#2#3{{\it Phys. Rept.} {\bf #1} (#2) #3}

\def\atmp#1#2#3{{\it Adv. Theor. Math. Phys.} {\bf #1} (#2) #3}
\def\jhep#1#2#3{{\it J. High Energy Phys.} {\bf #1} (#2) #3}

%
%

\lref\juan{J. M. Maldacena, ``The large $N$ limit of superconformal field
theories and supergravity", hep-th/9711200, \atmp{2}{1998}{231}.}

\lref\gkp{S. S. Gubser, I. R. Klebanov and A. M. Polyakov, 
``Gauge theory correlators
from non-critical string theory", hep-th/9802109,
\plb{428}{1998}{105}.}

\lref\wittenads{E. Witten, ``Anti-de-Sitter space and holography",
hep-th/9802150, \atmp{2}{1998}{253}.}

\lref\magoo{O. Aharony, S. S. Gubser, J. M. Maldacena, H. Ooguri and
Y. Oz, ``Large $N$ field theories, string theory and gravity,''
hep-th/9905111, \prep{323}{2000}{183}.}

\lref\polstr{J. Polchinski and M. J. Strassler, ``The string dual of a
confining four-dimensional gauge theory,'' hep-th/0003136.}

\lref\vafwit{C.~Vafa and E.~Witten, ``A strong coupling test of S duality,''
hep-th/9408074, \np{431}{1994}{3}.}

\lref\donwit{R.~Donagi and E.~Witten, ``Supersymmetric Yang-Mills
theory and integrable systems,'' hep-th/9510101, \np{460}{1996}{299}.}

\lref\gppz{L.~Girardello, M.~Petrini, M.~Porrati and A.~Zaffaroni,
``The supergravity dual of $\cn = 1$ super Yang-Mills theory,''
hep-th/9909047.}

\lref\dorey{N.~Dorey,
``An elliptic superpotential for softly broken $\cn = 4$ 
supersymmetric  Yang-Mills theory,'' hep-th/9906011,
\jhep{9907}{1999}{021}.}

\lref\frewit{D.~S.~Freed and E.~Witten,
``Anomalies in string theory with D-branes,''
hep-th/9907189.}

\lref\kacsmi{V.~G.~Kac and A.~V.~Smilga,
``Normalized vacuum states in $\cn = 4$ 
supersymmetric Yang-Mills quantum  mechanics with any gauge group,''
hep-th/9908096, \np{571}{2000}{515}.}

\lref\wittenbb{E.~Witten,
``Baryons and branes in anti de Sitter space,'' hep-th/9805112,
\jhep{9807}{1998}{006}.}

\lref\myers{R.~C.~Myers,
``Dielectric-branes,'' hep-th/9910053, \jhep{9912}{1999}{022}.}

\lref\hkr{A.~Hanany, B.~Kol and A.~Rajaraman,
``Orientifold points in M theory,'' hep-th/9909028,
\jhep{9910}{1999}{027}.}

\lref\bena{I. Bena, ``The M-theory dual of a 3 dimensional theory with
reduced supersymmetry,'' hep-th/0004142.}

\lref\hankol{A.~Hanany and B.~Kol,
``On orientifolds, discrete torsion, branes and M theory,''
hep-th/0003025.}

%

\Title{\vbox{\baselineskip12pt
\hbox{hep-th/0004151}\hbox{RUNHETC-2000-15}}}
{\vbox{
{\centerline { String Theory Duals for Mass-deformed }}
\vskip .1in
{\centerline { $SO(N)$ and $USp(2N)$ $\cn=4$ SYM Theories  }}
  }}
\centerline{Ofer Aharony\foot{oferah@physics.rutgers.edu} and
Arvind Rajaraman\foot{arvindra@physics.rutgers.edu}}
\vskip.1in
\centerline{ {\it Department of Physics and Astronomy, Rutgers University,
Piscataway, NJ 08855, USA}}
\vskip.1in
\vskip .5in

\centerline{\bf Abstract}

We generalize the results of Polchinski and Strassler regarding the
$\cn=1$-preserving mass deformation of $\cn=4$ $SU(N)$ SYM theories and its
string theory dual to $SO(N)$ and $USp(2N)$ gauge groups. The string theory
duals involve 5-branes wrapped on $\RP^2$. In order to match with the
field theory classification of vacua, the 3-brane charge carried by
such 5-branes must be shifted by a half, and this follows from a
conjectured generalization of results of Freed and Witten. Our results
provide an elegant physical picture for the classification of
classically massive vacua in the mass-deformed $\cn=4$ theories.

\Date{}

%

\newsec{Introduction}

The AdS/CFT correspondence \refs{\juan,\gkp,\wittenads} (see \magoo\
for a review) is
conjectured to be an exact duality between certain field theories and
certain compactifications of string/M theory. In particular, the
$\cn=4$ four dimensional SYM theory with $SU(N)$ gauge group is dual
to type IIB string theory on $AdS_5\times \bS^5$, and the $\cn=4$
theories with $SO(N)$ and $USp(2N)$ gauge groups are dual to type IIB
string theory on $AdS_5\times \RP^5$ \wittenbb. 

The correspondence was
originally stated for conformal field theories, but it can easily be
generalized also to relevant deformations of the conformal field
theories, which are realized as solutions of string/M theory with
particular boundary conditions depending on the deformation
parameter. Some string theory solutions corresponding to the mass
deformation of the $SU(N)$ $\cn=4$ theory were recently discussed in 
\polstr \foot{Very recently, these 
results were generalized to deformations of
the $d=3$ $\cn=8$ SCFTs \bena.}. The vacua of the
mass-deformed theory were described
in terms of configurations including 5-branes of
various types wrapped on $\IR^4\times \bS^2$. In particular, for the
mass deformation which preserves $\cn=1$ supersymmetry (which was
called the $\cn=1^*$ theory in \polstr), a one-to-one mapping was
found between
classical vacua of the field theory and configurations involving
purely D5-branes (though the supergravity approximation is only good
for some of these configurations, and others may be better described
in terms of other types of branes).

In this paper we wish to generalize the results of \polstr\ to the
case of the $\cn=1^*$ $SO(N)$ and $USp(2N)$ gauge theories. We will
find that again many of the classical vacua of the field theory have a
description in string theory in terms of configurations of D5-branes.
In particular, all the classically massive vacua (classified in
\kacsmi) have such a description. This gives a nice physical
realization of the mathematical results of \kacsmi\ (a different
realization of the same results was found in \hkr). 
The 5-branes in our configurations 
will be wrapped around $\IR^4\times \RP^2$, and in order to match with
the classical field theory results we will have to use a conjectured
generalization of the results of \frewit\ concerning charge
quantization to orientifolded backgrounds. This leads to a
half-integral shift in the 3-brane charge carried by 5-branes wrapped
on $\RP^2$, such that in the $USp(2N)$ case such D5-branes carry
integer 3-brane charges, while in the $SO(N)$ case the 3-brane charge
is an integer plus a half. 

The string theory solutions we find are almost
identical to those found in \polstr, and we will focus here just on
the aspects which are different in the $SO(N)$ and $USp(2N)$
theories. One such aspect is the fact that not all the classical vacua
may be mapped to D5-brane configurations. It would be interesting to
check if the other vacua have a dual description in terms of some
other configuration of 5-branes.

\newsec{The Vacua of the Mass-deformed $\cn=4$ SYM Theory}

The $\cn=4$ SYM theory may be described as an $\cn=1$ gauge theory
with three chiral superfields $\Phi_i$ ($i=1,2,3$) in the adjoint
representation, and a superpotential of the form 
\eqn\suppotzero{W = {{2\sqrt{2}}\over g_{YM}^2} \tr([\Phi_1, \Phi_2]
\Phi_3).}
There is one possible relevant deformation of this theory which
preserves $\cn=1$ supersymmetry, which is a superpotential of the form
\eqn\suppotone{\Delta W = {1\over g_{YM}^2} (m_1 \tr(\Phi_1^2) + m_2
\tr(\Phi_2^2) + m_3 \tr(\Phi_3^2)).}
We will take the masses to be equal, $m_1 = m_2 = m_3 = m$. It is
straightforward to generalize our results to the case of non-equal masses.

The classical vacua of the mass-deformed $\cn=1^*$
field theory are solutions to the F-term
and D-term equations,
\eqn\classeqs{\eqalign{
[\Phi_i, \Phi_j] &= -{m\over \sqrt{2}} \epsilon_{ijk} \Phi_k \cr
\sum_{i=1}^3 [\Phi_i, \Phi_i^\dagger] &= 0, \cr}}
up to gauge identifications. Setting the D-terms to zero and dividing
by the gauge group is the same as dividing by the complexified gauge
group $G_C$. Up to a constant, the F-terms give precisely the
commutation relation of the $SU(2)$ Lie algebra. Thus, as discussed in
\refs{\vafwit,\donwit}, the classical vacua are in one to one
correspondence with complex conjugacy classes of homomorphisms of the
$SU(2)$ Lie algebra to that of the gauge group $G$. Since such
homomorphisms have no infinitesimal deformations, the chiral
superfields are all massive in such vacua. The homomorphism generally
breaks the gauge group $G$ to a subgroup $H$, and classically the
gauge bosons of $H$ will be massless. Classically massive vacua
correspond to homomorphisms which break the gauge group $G$ completely
(or, at most, preserve a discrete subgroup).

For $G=SU(N)$, such homomorphisms are equivalent to $N$ dimensional
representations of $SU(2)$. Thus, there is one classical vacuum for
every partition of $N$, of the form $N = \sum_{i=1}^r n_i$ with positive
integers $n_i$. The $N\times N$ matrices corresponding to a particular
partition may be written in a block diagonal form, with blocks of size
$n_i\times n_i$ corresponding to the $n_i$-dimensional irreducible
representation of $SU(2)$. The gauge group is completely broken in the vacuum
corresponding to the $N$ dimensional representation, while in other
vacua there is (classically) an unbroken gauge group. If we denote the
number of times the $d$-dimensional representation appears in the
partition by $k_d$, the unbroken gauge group is $(\prod_{d=1}^N
U(k_d)) / U(1)$; the non-Abelian factors in this gauge group rotate
blocks which have the same size. Vacua containing unbroken non-Abelian
gauge groups are expected to split into several different vacua in the
quantum theory. If there are no unbroken Abelian gauge groups such
vacua will have a mass gap.

For other gauge groups, we are not aware of a complete classification
of the possible homomorphisms. However, the homomorphisms which
completely break the gauge group (up to possible discrete factors)
were classified in \kacsmi. For $G=SO(N)$ one such classically massive
vacuum was found for every partition of $N$ into distinct odd
integers, while for $G=USp(2N)$ one such vacuum was found for every
partition of $2N$ into distinct even integers\foot{This is related to the
fact that the odd-dimensional representations of $SU(2)$ are real,
while the even-dimensional representations are pseudo-real.}. As above,
these partitions may also be mapped into a block-diagonal form
\kacsmi. Partitions in which the (odd or even)
integers are not all distinct may
similarly be mapped to field theory vacua corresponding to 
block-diagonal matrices,
in which the gauge group is not completely broken; if there are $k$
blocks of the same size, the unbroken gauge group includes an $SO(k)$
factor. Again, vacua with unbroken non-Abelian gauge groups will split
in the full quantum theory. The vacua of the quantum theory (for any
gauge group) may be found by a generalization of the analysis of \dorey.

Unlike the $SU(N)$ case, it seems that for $SO(N)$ and $USp(2N)$ gauge groups
not all the solutions to \classeqs\ are given by the block-diagonal
solutions described above. In the $USp(2N)$ case, even the trivial
solution $\Phi_i=0$ is not included in our classification
above\foot{As we will see in the next section, this means that it 
is not dual to a background involving purely D5-branes.}.
Another example is $SO(6) \simeq SU(4)$ which has a solution
breaking the gauge group to $SU(2)\times U(1)$ which is not in our
classification. We will not attempt a complete classification of the
solutions to \classeqs\ here.

\newsec{The String Theory Description of the Vacua}

The $\cn=4$ SYM theory with gauge group $G=SU(N)$ is believed to be
dual to type IIB string theory compactified on $AdS_5\times \bS^5$ with
$N$ units of 5-form flux on the $\bS^5$
\refs{\juan,\gkp,\wittenads}. Similarly, the theories with gauge
groups $SO(N)$ and $USp(2N)$ are believed to be dual to
compactifications of type IIB string theory on $AdS_5\times \RP^5$,
in which non-orientable string worldsheets wrapping the non-trivial
$\RP^2 \subset \RP^5$ are allowed. The $SO(N)$ and $USp(2N)$ cases are
distinguished by discrete torsion. The RR
and NS-NS 3-form fields are twisted by the orientifolding, therefore
their cohomology is classified by $H^3(\RP^5, \tilde{\IZ}) =
\IZ_2$, as explained in \wittenbb. 
Thus, for each 3-form field there are two choices of the
discrete torsion, leading to four different theories. The theory with
no torsion is $SL(2,\IZ)$ invariant and is dual to the $SO(2N)$
gauge theory. The theory with only RR torsion is dual to an $SO(2N+1)$
gauge theory, while the two theories with NS-NS torsion are dual to
two $USp(2N)$ theories which differ in their soliton spectrum \hankol.

The AdS/CFT correspondence may be extended to include relevant
deformations of the conformal field theory, which are mapped to string
theory in asymptotically AdS backgrounds with particular boundary
conditions corresponding to the deformation. In particular, the mass
deformation we analyzed above (for equal masses) corresponds to a
background including 3-form field strengths (as reviewed in section
4.3 of \magoo). We expect to have a string theory solution with the
appropriate boundary conditions corresponding to every vacuum of the
quantum field theory after the mass deformation.

Recently, Polchinski and Strassler \polstr\ gave a beautiful physical
picture of the vacua of the mass-deformed $SU(N)$ theory, in the limit
where supergravity is a good approximation. They noted, following
\myers, that $n$ 3-branes carrying 3 scalar fields whose VEVs are an
$n$-dimensional representation of $SU(2)$ (as described above) may be
thought of as a D5-brane wrapped on an $\bS^2$ (in the directions
corresponding to the 3 scalar fields), carrying $n$ units of D3-brane
charge coming from a $U(1)$ gauge field carrying $n$ units of flux on
the $\bS^2$. Thus, they interpreted the classical field theory vacua we
described above, corresponding to a partition $N=\sum_{i=1}^r n_i$, as
being dual to string theory configurations including $r$ D5-branes,
each carrying $n_i$ units of 3-brane charge. The geometry of the
5-branes is $\IR^4\times \bS^2$, where the $\bS^2$ is an equator in the
$\bS^5$ of $AdS_5\times \bS^5$, and the $\IR^4$ corresponds to a constant
radial position in $AdS_5$ (in Poincar\'e coordinates). 

Polchinski and Strassler constructed
approximate string theory solutions corresponding to these vacua
\polstr, and
found that the $i$'th D5-brane sits at a radial position in AdS space
which is proportional to $n_i$. This reproduces elegantly the
classically unbroken gauge group described above, which is just the
product of the gauge groups on the different D5-branes (which is
enhanced to a non-Abelian group when the D5-branes overlap), up to an
overall $U(1)$ factor which may be gauged away \polstr. The
supergravity approximation used in \polstr\ is only valid if all the
$n_i$ are much bigger than $\sqrt{g_s N} \gg 1$; if this does not hold
there are sometimes alternative descriptions of the same vacua in
terms of other types of 5-branes \polstr. We will ignore this issue
here and assume that in some sense the D5-brane vacua still exist even
if the supergravity approximation breaks down.

Most of the results of \polstr\ apply also to the $SO(N)$ and
$USp(2N)$ theories. The supergravity fields on $AdS_5\times \RP^5$ are
a $\IZ_2$ projection of the fields on $AdS_5\times \bS^5$, and this
projection preserves the fields that are excited in the solutions of
\polstr. Hence, in the supergravity approximation we can find the same
solutions involving 5-brane shells as in the $SU(N)$ case of
\polstr. The results of \polstr\ concerning confinement, screening,
baryons, condensates, domain walls, instantons, flux tubes, and so on, 
have obvious generalizations to the $SO(N)$ and $USp(2N)$ cases.
The 5-branes now wrap an $\RP^2$ which is an equator of $\RP^5$
instead of wrapping an $\bS^2$. As discussed in \wittenbb, such a
wrapping is consistent with the orientifold projection since it
inverts the 3-form field coupling to the 5-brane charge.

The main difference between wrapping 5-branes on $\bS^2$ and on $\RP^2$
is in the allowed amounts of 3-brane charge the 5-brane can carry. The
3-brane charge arises from a term in the effective action of the
D5-brane, of the form 
\eqn\term{\int d^6x\ C^{(4)} \wedge (F-B_{NS}).} 
Thus, the
3-brane charge is the integral of $(F-B_{NS})$ over the compact
2-cycle. For an $\bS^2$ in $AdS_5\times \bS^5$, this integral gives an
integer 3-brane charge by the usual quantization condition. However,
in general backgrounds an analysis of worldsheet global anomalies
can lead to a shift in this quantization \frewit. The quantization in
orientable backgrounds, for a D-brane worldvolume of the form
$\IR^{p-1}\times Q$ (with a 2-cycle $Q$) was found to be
\eqn\charges{\int_Q {F \over 2\pi} = {w_2(Q) \over 2}\ (mod\ 1),}
where $w_2(Q)$ is the second Stiefel-Whitney class, normalized to take
values in $\{0,1\}$. Thus, if $Q$ has a non-trivial $w_2$, the
contribution from the gauge field to the 3-brane charge will be
an integer plus a half. The results of \frewit\ have not yet been
generalized to include non-orientable backgrounds. However, it seems
that the formula \charges\ still holds in these backgrounds
\refs{\wittenbb,\frewit}, even though $F$ has to be generalized to a
twisted gauge bundle if $Q$ is non-orientable. We will conjecture that
this formula holds for $Q=\RP^2$, and the matching with the gauge
theory results described in the previous section will give strong
evidence for this.

For our case of a 5-brane on $\IR^4\times \RP^2$, the second
Stiefel-Whitney class is non-trivial, since for 2-cycles $w_2(Q) =
e(Q)\ (mod\ 2)$ where $e(Q)$ is the Euler class, which equals one for
$\RP^2$. Thus, the gauge field contribution to the 3-brane charge is an
integer plus one half in this case. An additional contribution can
come from the NS-NS 2-form field $B_{NS}$, integrated over $\RP^2$
($B_{NS}$ is twisted under the orientifolding so we can integrate it
over $\RP^2$). As discussed in \wittenbb, this integral is exactly the
discrete torsion of the NS-NS 3-form field strength, which is
classified by $H^3(\RP^5, \tilde{\IZ}) = \IZ_2$. As discussed above, 
for the backgrounds
corresponding to $SO(2N)$ and $SO(2N+1)$ gauge groups, this discrete
torsion vanishes, while for $USp(2N)$ it gives 
\eqn\bfield{\int_{\RP^2} {B_{NS}\over 2\pi}
= {1\over 2}\ (mod\ 1).} 
Thus, we find that for $SO(2N)$ and $SO(2N+1)$
theories, the 5-branes carry 3-brane charges in $\IZ+{1\over 2}$,
while for $USp(2N)$ groups they carry integer 3-brane charges.

In the $SU(N)$ case, we mapped the classical field theory vacua to
string theory configurations where all of the 3-brane flux was carried
by D5-branes, leaving flat space at the origin \polstr. In our case,
the total 3-brane flux cannot all be carried by D5-branes. One way to
see this is from the fact that the total flux is not an integer, or
even a half-integer, but rather is given by 
\eqn\gfive{\int_{\RP^5} {G^{(5)}\over
2\pi} = \cases{N-{1\over 4} &for $SO(2N)$,\cr\noalign{\vskip 2pt} 
                      N+{1\over 4} &for $SO(2N+1)$,\cr\noalign{\vskip 2pt} 
                      N+{1\over 4} &for $USp(2N)$;\cr} }
these charges may be deduced from the
charges of D3-branes stuck on orientifold planes before taking the
near-horizon limit. Thus, at the origin we always remain with some
charge that must be carried by an orientifold. Equivalently, it is
clear geometrically that since the transverse space is orientifolded,
we must remain with an orientifold at the origin and not just with
flat ten dimensional space\foot{In the $USp(2N)$ cases, the D5-branes
we added do not change the discrete torsion, so the orientifold we
remain with at the origin carries the same discrete 
torsion as the
                      original background. In the $SO(N)$ cases, each
                      D5-brane is actually a domain wall along which
                      the RR discrete torsion changes \wittenbb, 
so we could end
                      up at the origin with either sign for this
                      discrete torsion. The orientifold with the RR
                      discrete torsion may be viewed as the
                      orientifold without the torsion with a half
                      D3-brane stuck on top of it. It seems that such
                      a half D3-brane may be identified with a one
                      dimensional block in the scalar matrices (recall
                      that for small blocks the description in terms
                      of D5-branes breaks down). With this
                      identification it does not matter which type of
                      orientifold we end up with, and we will assume
                      below that we end up with no discrete torsion.}.

Now, we can list the string theory vacua where the 3-brane flux is
completely carried by D5-branes (except for the contribution of the
remaining orientifold), as in the vacua of \polstr\ described
above. For $SO(N)$ gauge groups this charge is $N/2$,
and this should be reproduced as the sum of contributions in
$\IZ+{1\over 2}$; hence, we find one such vacuum for every partition
of $N$ into odd integers. For $USp(2N)$ gauge groups, this charge is
$N$, which should arise as the sum of contributions in $\IZ$; thus, we
find one vacuum for every partition of $N$.

These results agree precisely with the gauge theory results described
above, if a single D5-brane wrapped around $\RP^2$ carries no gauge
group while multiple D5-branes carry some non-trivial gauge group;
then, we would find no remaining gauge groups exactly when the
partitions involve distinct integers. This fact is indeed true,
because the orientifolding projects out part of the four dimensional 
gauge fields
living on the D5-branes wrapped on $\RP^2$, leaving only an $SO(N)$
gauge group for $N$ D5-branes (instead of the original $U(N)$). 

Our description (and that of \polstr) of the field theory vacua in
terms of supergravity plus D5-branes is only valid when the D3-brane
charges $n_i$ carried by the D5-branes all satisfy $n_i \gg \sqrt{g_sN}
\gg 1$. However, the matching with the field theory analysis indicates
that these vacua make sense in string theory even when this condition
is not satisfied. As discussed in \polstr, in many cases where
supergravity breaks down there is an alternative description involving
other types of 5-branes, like NS 5-branes. This will be true also for
the $SO(N)$ and $USp(2N)$ theories. The allowed charges carried by
general $(p,q)$ 5-branes wrapped on $\RP^2$ may be derived by an
$SL(2,\IZ)$ transformation of the results we described above. For
example, a wrapped NS 5-brane will carry an integer 3-brane charge 
when the RR discrete torsion is non-zero, and the charge
will be shifted by a
half when there is no RR discrete torsion.

As discussed at the end of section 2, there are also classical field theory
vacua which do not correspond to partitions of the type we discussed
here, so we cannot map them to vacua including purely D5-branes. It
would be interesting to find string theory duals for these vacua.

\vskip1cm
\noindent{\bf Acknowledgements}

\noindent

We would like to thank D.-E. Diaconescu, M. R. Douglas,
G. Moore and M. Strassler for
useful discussions. This work was supported in part by DOE grant
DE-FG02-96ER40559.

\listrefs

\end